\documentclass[twocolumn,prl,aps,showpacs,superscriptaddress]{revtex4}

\usepackage{graphicx}
\usepackage{endnotes}
\usepackage{color}

\begin{document}
\newcommand{\fecosi}{Fe$_{1-x}$Co$_x$Si}
\newcommand{\ZrZn}{ZrZn$_2$}
\newcommand{\uge}{UGe$_2$}
\newcommand{\MnSi}{MnSi\,-\,110 }
\newcommand{\Mnsi}{MnSi\,-\,111 }
\newcommand{\Tc}{$T_{c}$ }
\newcommand{\rhoxx}{$\rho_{xx}$ }
\newcommand{\rhoxy}{$\rho_{xy}$ }

\renewcommand{\floatpagefraction}{0.5}

\title{Skyrmion Lattice in a Doped Semiconductor}

\author{W. M\"unzer}
\affiliation{Physik Department E21, Technische Universit\"at M\"unchen,
D-85748 Garching, Germany}

\author{A. Neubauer}
\affiliation{Physik Department E21, Technische Universit\"at M\"unchen,
D-85748 Garching, Germany}

\author{S. M\"uhlbauer}
\affiliation{Physik Department E21, Technische Universit\"at M\"unchen,
D-85748 Garching, Germany}

\author{C. Franz}
\affiliation{Physik Department E21, Technische Universit\"at M\"unchen,
D-85748 Garching, Germany}

\author{T. Adams}
\affiliation{Physik Department E21, Technische Universit\"at M\"unchen,
D-85748 Garching, Germany}

\author{F. Jonietz}
\affiliation{Physik Department E21, Technische Universit\"at M\"unchen,
D-85748 Garching, Germany}

\author{R. Georgii}
\affiliation{Forschungsneutronenquelle Heinz-Maier-Leibnitz (FRM II),
Technische Universit\"at M\"unchen, D-85748 Garching, Germany}
\affiliation{Physik Department E21, Technische Universit\"at M\"unchen,
D-85748 Garching, Germany}

\author{P. B\"oni}
\affiliation{Physik Department E21, Technische Universit\"at M\"unchen,
D-85748 Garching, Germany}

\author{B. Pedersen}
\affiliation{Forschungsneutronenquelle Heinz-Maier-Leibnitz (FRM II),
Technische Universit\"at M\"unchen, D-85748 Garching, Germany}

\author{M. Schmidt}
\affiliation{MPI f\"ur chemische Physik fester Stoffe, N\"othnitzer Stra§e 40, D-01187 Dresden, Germany}

\author{A. Rosch}
\affiliation{Institut f\"ur Theoretische Physik, Universit\"at zu K\"oln, Z\"ulpicher Str. 77, D-50937 Cologn, Germany}

\author{C. Pfleiderer}
\affiliation{Physik Department E21, Technische Universit\"at M\"unchen,
D-85748 Garching, Germany}

\date{\today}

\begin{abstract}
We report a comprehensive small angle neutron scattering study (SANS) of the magnetic phase diagram of the doped semiconductor {\fecosi} for $x=0.2$ and 0.25. For magnetic field parallel to the neutron beam we observe a six-fold intensity pattern under field-cooling, which identifies the A-phase of {\fecosi} as a skyrmion lattice. The regime of the skyrmion lattice is highly hysteretic and extents over a wide temperature range, consistent with the site disorder of the Fe and Co atoms. Our study identifies {\fecosi} is a second material after MnSi in which a skyrmion lattice forms and establishes that skyrmion lattices may also occur in strongly doped semiconductors.  
\end{abstract}

\pacs{72.80.Ga, 72.15.-v, 72.25.-b, 75.30.-m}

\vskip2pc

\maketitle

Recently a skyrmion lattice was identified in the cubic B20 system MnSi \cite{mueh09,neub09}, that is, magnetic order representing a crystallization of topologically stable, particle-like knots of the spin structure originally anticipated to occur in anisotropic materials \cite{bogd89}. This raises the question for further magnetic materials with skyrmion lattices and if they are a general phenomenon in cubic magnets without inversion symmetry as suggested by our theoretical treatment in \cite{mueh09}. Because MnSi is a pure metal, an additional question concerns if skyrmion lattices are sensitive to disorder and whether they also exist in semiconductors and insulators. More generally, the microscopic identification of a skyrmion lattice in MnSi represents also a showcase for similar lattice structures considered in nuclear physics \cite{skyr62,kleb85}, quantum Hall systems \cite{sond93,brey95}, and liquid crystals \cite{bogd03}. 

The B20 transition metal silicides TSi (T=Fe,Co,Mn) are ideally suited to clarify these questions. While FeSi is a nonmagnetic insulator with strong electronic correlations \cite{aepp92} its sibling CoSi is a diamagnetic metal \cite{shin66}. For increasing Co-content the series {\fecosi} displays a insulator to metal transition at $x=0.02$ becoming increasingly metallic with a dome of helimagnetic order in the range $0.05\leq x\leq0.7$ \cite{beil83}. Being a strongly doped Si-based semiconductor the magneto-transport properties of {\fecosi} attract great interest  \cite{many00,onos05}. 

The helimagnetic order in {\fecosi} results from a hierarchy of energy scales, of which ferromagnetic exchange on the strongest scale favors parallel spin alignment and Dzyaloshinsky-Moriya spin-orbit interactions, permitted in the non-centrosymmetric B20 structure, favor perpendicular spin alignment on a weaker scale. With increasing $x$ the wavelength of the resulting helical modulation varies from about 200\,{\AA} to 2000\,{\AA} \cite{beil83,grig07a}. The propagation direction of the helical modulation is determined by crystal-field interactions on the weakest scale. Based on SANS experiments \cite{ishi95,grig07a,take09} and Lorentz force microscopy \cite{uchi06} it was concluded that the propagation axis of the helix in {\fecosi} is $\langle100\rangle$ for all $x$. While this differs distinctly from the propagation direction in MnSi being $\langle111\rangle$, it is perfectly consistent with the cubic crystal structure \cite{bak80,naka80}. Further, while the DM interaction and the crystal structure have the same chirality in MnSi \cite{tana85}, they have opposite chirality in {\fecosi} \cite{grig09}. Finally, due to the strong doping in {\fecosi} there must be also site disorder of the Fe and Co atoms even in excellent single crystals. Thus, although {\fecosi} and MnSi exhibit rather similar magnetic properties, there are strong differences of the underlying electronic structure.

The magnetic phase diagram of {\fecosi} exhibits three prominent features \cite{ishi95,grig07a,take09}. First, a state with a helimagnetic modulation up to a critical field $B_{c1}$, referred to in the following as zero-field cooled (ZFC) state. Second, for a magnetic field in the range $B_{c1}<B<B_{c2}$ a helical modulation parallel to the magnetic field is stabilized, forming the so-called conical state. Third, in the so-called A-phase, a small pocket just below $T_c$, the helical modulation is perpendicular to the applied magnetic field. Prior to the work reported here all studies were carried out with the field perpendicular to the neutron beam. Because this configuration is mostly sensitive to helical components parallel to the field  and only selected helical components perpendicular to the field the information on the A-phase was incomplete and it was believed that the A-phase represents a single-$Q$ helical state \cite{ishi95,grig07a,take09}.

Here we report a comprehensive SANS study of the magnetic phase diagram of {\fecosi} for $x=0.2$ and $x=0.25$. In contrast to previous SANS studies \cite{ishi95,grig07a,take09} we explore the magnetic phase diagram also with the magnetic field \textit{parallel} to the incident neutron beam, a configuration that is sensitive to \textit{all} helical modulations perpendicular to the field.  We also determined differences of the phase diagram for zero-field cooling (ZFC) and field-cooling (FC). As our main result we identify the A-phase of {\fecosi} as a hexagonal skyrmion lattice akin that seen in MnSi. By comparison to MnSi, the skyrmion lattice yields hysteretic features and may be observed over a large temperature range. Thus we identify a second example of a skyrmion lattice, however, in a doped semiconductor with strong disorder. In addition we find that the ZFC state, long known to be metastable \cite{ishi95}, is characterized by scattering intensity on the surface of a small sphere in reciprocal space with broad maxima in the $\langle 110\rangle$ direction, reminiscent of the partial magnetic order in MnSi at high pressure \cite{pfle04}. 

\begin{figure}
\includegraphics[width=0.75\linewidth,clip=]{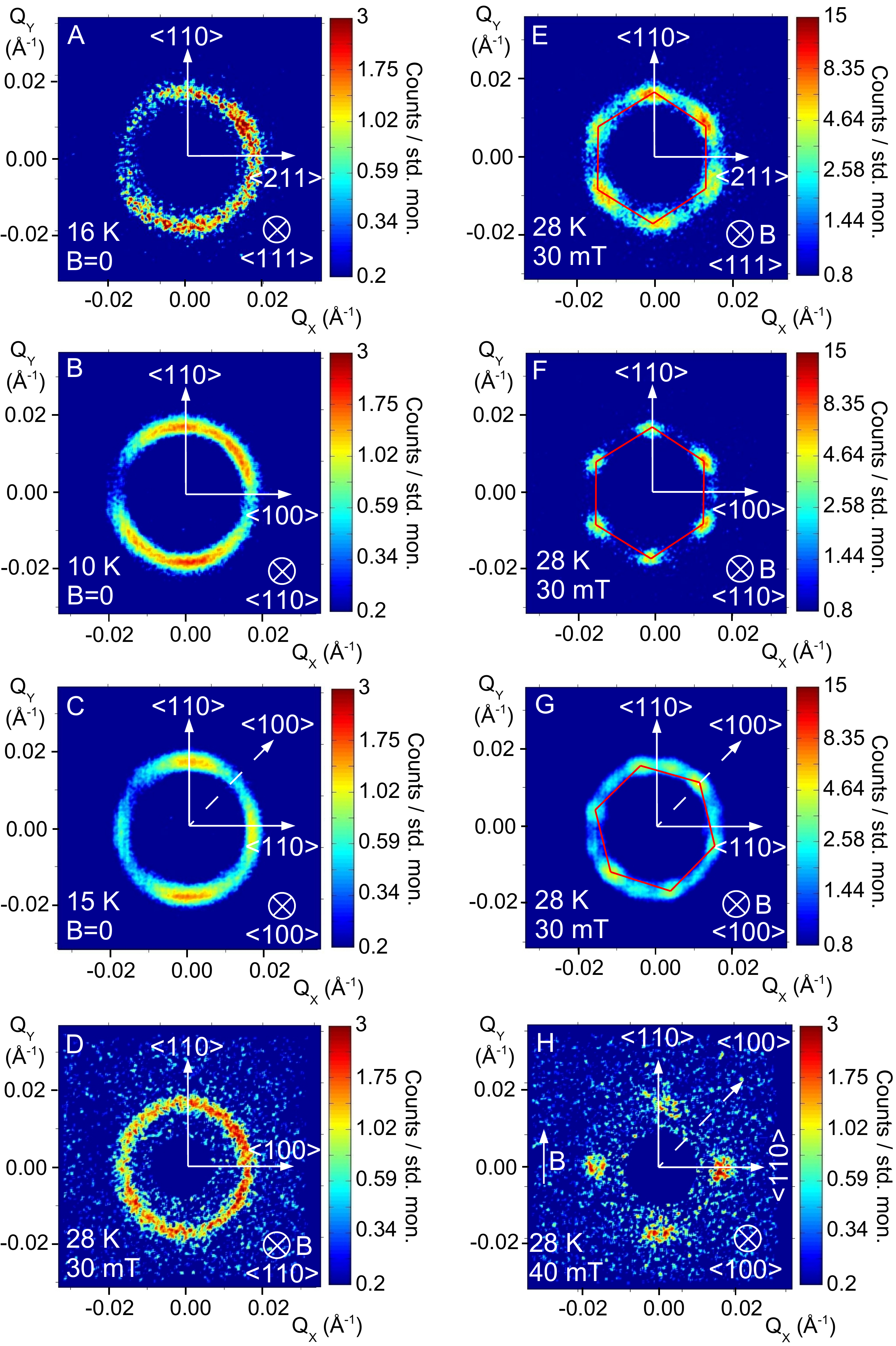}
\caption{Typical SANS data of {\fecosi} for $x=0.2$. (A) through (C) Intensity after zero field cooling (ZFC). (D) A-phase after ZFC for increasing $B$ parallel to the neutron beam. (E) through (G) Six-fold intensity distribution in the A-phase after field-cooling for $B$ parallel to the neutron beam. (H) A-phase after field-cooling for $B$ perpendicular to the neutron beam. In all panels the intensity on the left-hand side is reduced due to a small inefficiency of the neutron detector.}
\label{figure1}
\end{figure}

For our study several single crystals were grown at TUM by optical float-zoning with a UHV-compatible image furnace \cite{neub09a}. Laue x-ray diffraction, EDX, polarized light microscopy and single-crystal neutron diffraction established an excellent sample quality. The magnetotransport properties, magnetization and specific heat were in excellent agreement with the literature \cite{onos05} as well as small single-crystals grown by vapor transport at MPI-CPfS, Dresden.  All samples were carefully oriented using Laue x-ray diffraction. A large single crystal with $x=0.2$, studied most extensively, was additionally oriented on the diffractometer RESI at FRM II. Using neutrons with a wavelength $\lambda=1.0402\,{\rm \AA}\pm1\%$ we determined ten Bragg peaks that were fitted simultaneously. The sample had a lattice constant $a=4.483\pm0.02\,{\rm \AA}$ with an excellent resolution-limited mosaic spread better than $0.2^{\circ}$. We further confirmed within the resolution limit that there were no superstructures,  structural short-range order, or lattice distortions.  The modulus of the helical modulation $Q\approx0.017\,{\rm \AA^{-1}}$, transition temperature $T_c\approx30\,{\rm K}$ and the lattice constant were in excellent agreement with Refs.\cite{shim90,ishi95,take09}, but differed from crystals grown in a tri-arc furnace \cite{grig07a}.

Our SANS studies were carried out at the diffractometer MIRA at FRM II, using neutrons with a wavelength $\lambda=9.7\,{\rm \AA}\pm 5\%$. Backgrounds were determined at high temperatures and subtracted accordingly. A Cd marker on the sample support confirmed that the sample was always oriented correctly. Unfortunately the $^3$He delayline detector displayed  an inefficiency on the left-hand side, consistently visible in Fig.\,\ref{figure1}. This does not affect the conclusions of the work reported here. The sample was cooled with a pulse-tube cooler. Magnetic fields up to 0.3\,T were applied with a bespoke set of Helmholtz-coils. Data were recorded for a fixed sample orientation following rocking scans with respect to a vertical axis of typically $\pm15^{\circ}$. 

In the following we focus on the ZFC state and the A-phase. Data recorded in the conical state are not discussed in detail since they agree very well with previous studies. To determine the three-dimensional distribution of scattering intensity of the ZFC state we have recorded data for different sample orientations. As shown in Fig.\,\ref{figure1}\,A, B and C for $\langle111\rangle$, $\langle110\rangle$ and $\langle100\rangle$, the ZFC intensity for $x=0.2$ is characterized by a uniform ring, a ring with minima for $\langle100\rangle$ (horizontal direction), and a ring with minima for $\langle100\rangle$ (diagonals), respectively. Taking together Figs.\,\ref{figure1}\,A, B and C, the ZFC state is characterized by intensity on the surface of a small sphere in reciprocal space with broad intensity maxima for $\langle110\rangle$ \cite{rotate}. This contrasts previous studies, where a similar ZFC intensity distribution was reported with maxima for $\langle100\rangle$ \cite{ishi95,grig07a,take09,zfc}. We have therefore confirmed our results for several samples with $x=0.2$ from different growths. Moreover, for $x=0.25$ we find a similar scattering distribution in the ZFC state (data not shown). Thus data in Ref.\,\cite{grig07a,take09} may have either been indexed incorrectly or an unexplained sample dependence occurs for $x=0.2$ and 0.25, where our samples by all accounts seem excellent. The broad intensity maxima for $\langle110\rangle$ in the ZFC state are also unusual, because the propagation direction of the helical modulation in the cubic B20 crystal structure in leading order may be either parallel to $\langle111\rangle$ or $\langle100\rangle$ \cite{bak80,naka80}. In fact, the ZFC intensity distribution corresponds to the partially ordered state of MnSi \cite{pfle04}, which inspired proposals of spontaneously forming spin structures such as skyrmion lattices \cite{binz06,roes06,tewa06,fisc08}. 

\begin{figure}
\includegraphics[width=0.75\linewidth]{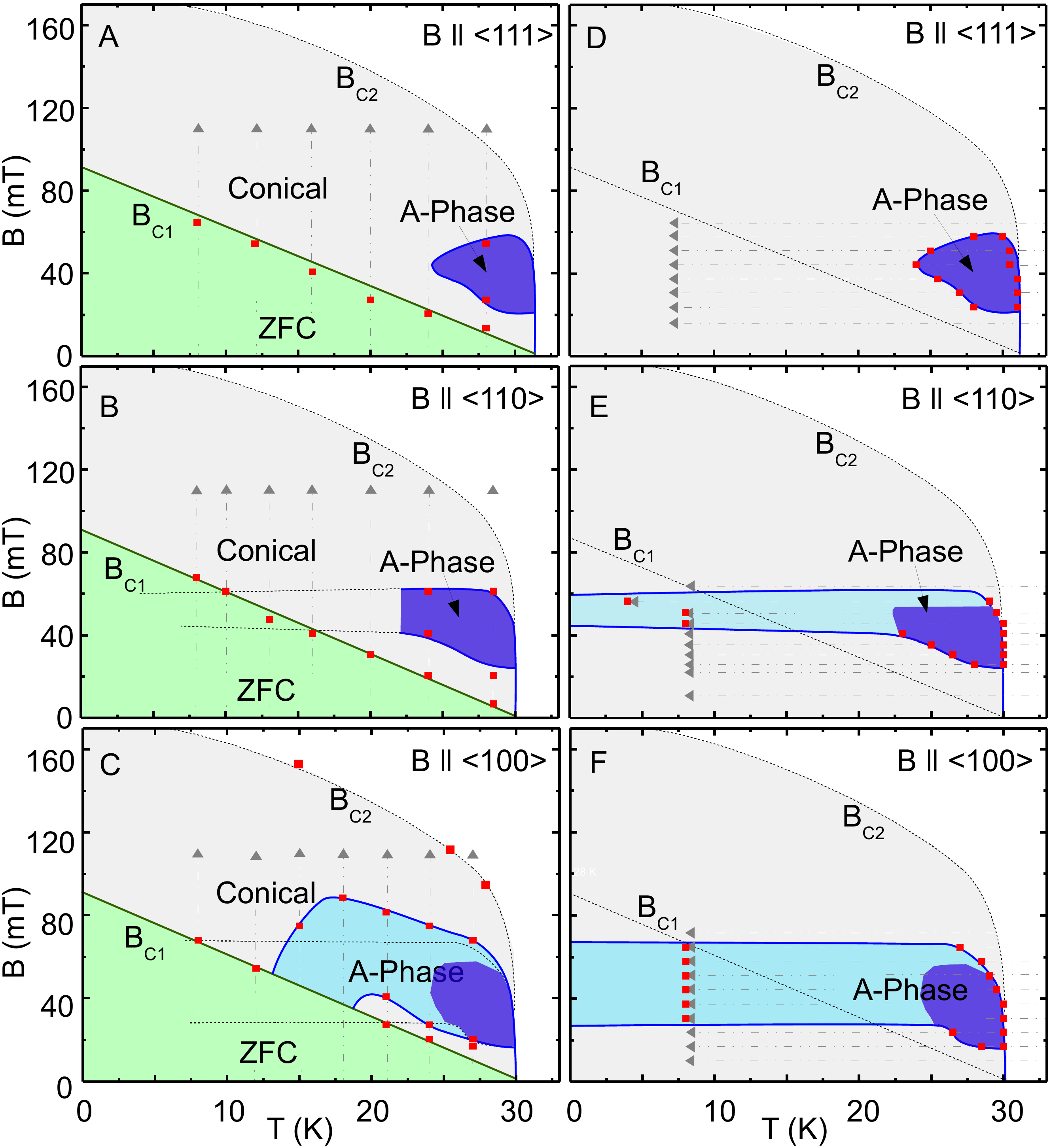}
\caption{Magnetic phase diagram of {\fecosi} for $x=0.2$. Gray arrows show the location of temperature and field scans. Red data points show transition fields and temperatures as determined in these scans. Panel (A) through (C) show the magnetic phase diagrams after ZFC; in the green shaded area the scattering intensity of the ZFC state is observed. Panel (D) through (F) show the magnetic phase diagrams for FC. Across the dark blue area the intensity displays a maximum.}
\label{figure3}
\end{figure}

Shown in Fig.\,\ref{figure1}\,D through G are typical intensity patterns in the A-phase. These patterns are exclusively seen in the plane perpendicular to the magnetic field. After ZFC and for increasing field we observe a uniform ring of intensity regardless of the crystallographic orientation (Fig.\,\ref{figure1}\,D). In contrast, for FC the intensity pattern in the A-phase is characterized by a six-fold symmetry. For the $\langle111\rangle$ and $\langle110\rangle$ directions six spots are observed as shown in Fig.\,\ref{figure1}E and F, respectively, where two of the spots are aligned with a $\langle110\rangle$ axis. Yet, for the $\langle100\rangle$ direction the six-fold pattern is composed of twelve spots comprising two sets of six spots, as shown in Fig.\,\ref{figure1}G. This is characteristic of two domain populations, where each domain yields the same hexagonal symmetry seen for $\langle111\rangle$ and $\langle110\rangle$. Interestingly, two spots of each domain population are now parallel to $\langle100\rangle$. We have finally also confirmed the formation of the conical phase, which coexists with the A-phase in Fig.\,\ref{figure1}\,H (horizontal and vertical spots are due to the A-phase and conical phase, respectively).

Shown in Fig.\,\ref{figure3} is the magnetic phase diagram for $x=0.2$ (similar diagrams for $x=0.25$ are not shown).  We find excellent agreement with published work where data are available. Starting from a ZFC state we observe: (i) scattering intensity with broad maxima for $\langle110\rangle$ up to $B_{c1}$, where $B_{c1}(T\to0)\approx95\,{\rm mT}$ and $dB_{c1}/dT\approx-3\,{\rm mT\,K^{-1}}$ are isotropic (green shading), (ii) a conical phase as reported before (gray shading), and (iii) a ring of intensity in the A-phase, where the wave vector is perpendicular to the field (blue shading). For FC the salient features are: (iv) the ZFC scattering intensity is never observed, (v) scattering intensity in the A-phase with a six-fold symmetry and wave vectors perpendicular to the field (blue shading), and (vi) a small field range in which the A-phase exists down to the lowest temperatures (this is not seen for field in the $\langle111\rangle$ direction). The regime where the intensity in the A-phase displays a maximum is shown as dark blue shading, while the remaining regime is shown in light blue shading.

\begin{figure}
\includegraphics[width=0.85\linewidth,clip=]{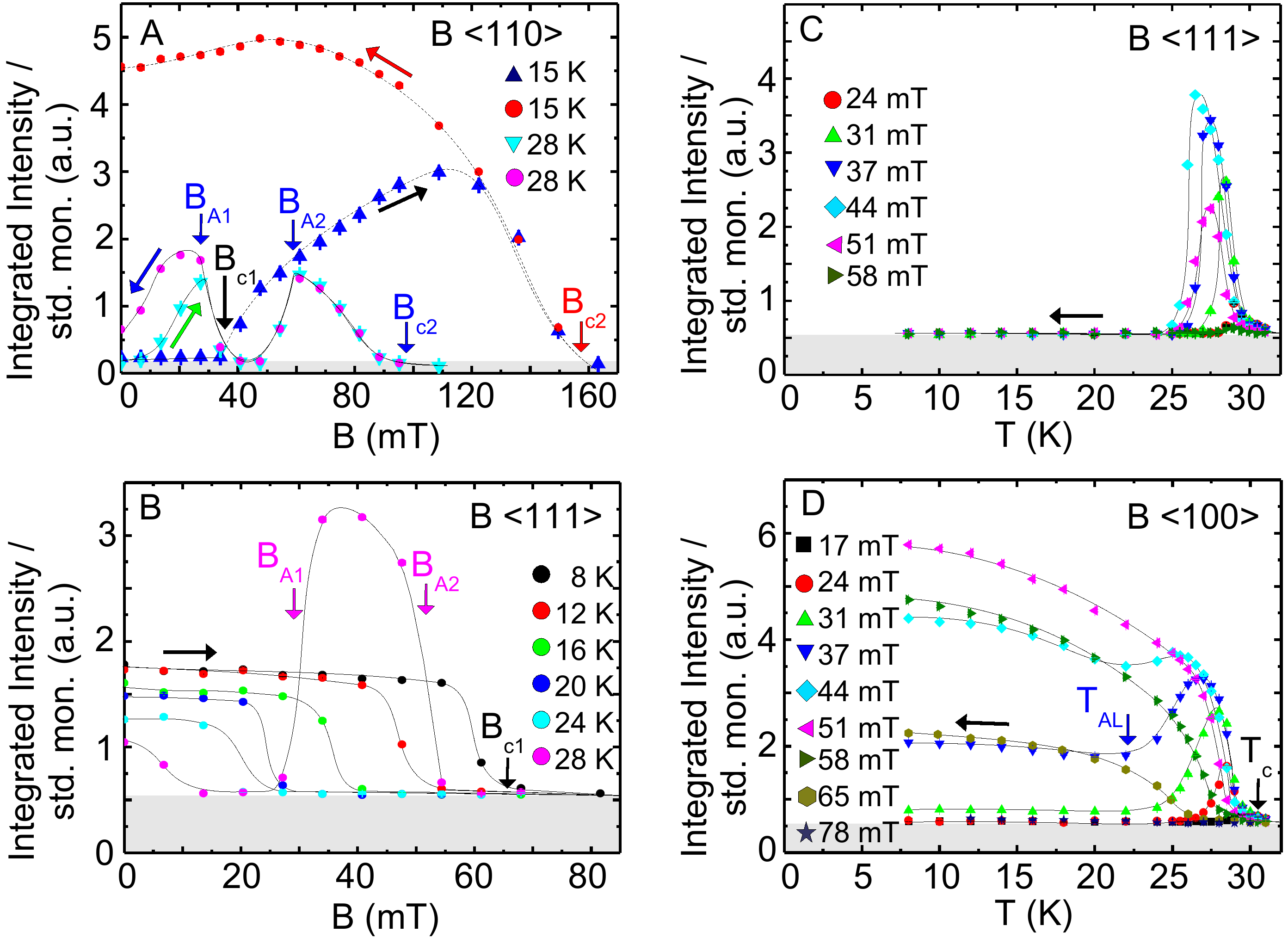}
\caption{Temperature and magnetic field dependences of the integrated scattering intensity. (A) Intensity for $B\parallel\langle110\rangle$ and $B$ perpendicular to the neutron beam. (B) Intensity for $B\parallel\langle111\rangle$ and $B$ parallel to the neutron beam. (C) and (D) Intensity for $B$ parallel to neutron beam under FC.  All panels use the same arbitrary scale; gray shading shows the background.}
\label{figure2}
\end{figure}

Metastable aspects of the ZFC state and the A-phase are best illustrated by the integrated intensity shown in Fig.\,\ref{figure2}. For field scans with $B$ \textit{perpendicular} to the neutron beam we reproduced published results \cite{grig07a}. As shown in Fig.\,\ref{figure2}\,A for $\langle110\rangle$, there is essentially no intensity up to $B_{c1}$ for increasing $B$ and low $T$ above which the conical phase appears. In contrast, for decreasing field the intensity remains unchanged high below $B_{c1}$. In the vicinity of $T_c$ the intensity displays a reversible minimum in the A-phase between $B_{A1}$ and $B_{A2}$, where the intensity includes contributions of the A-phase and the conical phase (see Fig.\,\ref{figure1}\,H). 

Scans for field \textit{parallel} to the neutron beam shown in Fig.\,\ref{figure2}\,B, C and D are the counterpart to that shown in Fig.\,\ref{figure2}\,A, notably they show \textit{all} helical components perpendicular to the field. For $\langle111\rangle$ and increasing field the ZFC intensity well below $T_c$  is unchanged until it vanishes above $B_{c1}$ (Fig.\,\ref{figure2}\,B). However, in field scans of decreasing strength the ZFC intensity is not recovered (data not shown).  Near $T_c$ the intensity reversibly shows a maximum in the A-phase (data at 28\,K).  In temperature scans a corresponding maximum is seen in the A-phase (see Fig.\,\ref{figure2}\,C  for $\langle111\rangle$). An additional feature emerges in temperature scans for $\langle 100 \rangle$ and $\langle 110 \rangle$, shown in Fig.\,\ref{figure2}\,D for $\langle 100 \rangle$.  For $B>31\,{\rm mT}$ and FC the intensity displays not only a maximum, but also an additional increase down to the lowest temperatures. The regime of the maximum near $T_c$ agrees between ZFC and FC for all three directions (compare dark blue shading in Fig.\,\ref{figure3}\,A, B and C with D, E and F, respectively). Interestingly, for $\langle100\rangle$ the A-phase even borders directly on the ZFC state (Fig.\,\ref{figure3}\,C).

The scattering pattern and hysteretic behavior reported here are fully consistent with a scenario where the A-phase is thermodynamically stable in the dark blue areas and metastable in the light blue region. The characterizing 6-fold pattern is, however, directly observable only for FC. Due to the disorder and the extremely weak anisotropy \cite{mueh09}, the spin structure is not able to lock to the underlying crystalline structure for ZFC so that only rings of intensity can be observed as shown in Fig.\ref{figure1}\,D. As shown in Ref.\,\cite{mueh09} the A-Phase can be identified as a hexagonal lattice of skyrmion tubes. The trend to align two intensity maxima along $\langle110\rangle$ for field in the $\langle 111 \rangle$ and $\langle 110 \rangle$ directions can be easily explained in the model described in Ref.\,\cite{mueh09} by adding, e.g., an anisotropy term $\sum_{\vec{k}} (k_x^6+k_y^6+k_z^6) \vec{m}_{\vec{k}}  \vec{m}_{\vec{-k}}$ with a positive coefficient. 

The observation of two domain populations for $\langle100\rangle$ (Fig.\,\ref{figure1}\,G), which change as a function of field and temperature, unambiguously distinguishes the skyrmion lattice from a single-$Q$ multi-domain state. However, by symmetry, the contribution of the 6th order anisotropy term discussed above (and similar terms with a $\pi/2$ rotation symmetry) vanishes to leading order for a field parallel $\langle 100 \rangle$. Indeed, the pattern shown in Fig.\ref{figure1}\,G differs qualitatively from Figs.\ref{figure1}\,E and \ref{figure1}\,F. As the Co atoms reduce the symmetry locally one may speculate that pinning terms of random sign arising from the disorder explain the two six-fold domain populations shown in Fig.\ref{figure1}\,G.

Our data further allude to a possible analogy of skyrmion lattices with superconducting flux line lattices \cite{bogd89}. For instance, the superconducting flux lattice in Nb shows two domain populations for $\langle100\rangle$ driven by the frustration between the sixfold symmetry of the flux lattice and the four-fold symmetry of crystal lattice (see e.g. \cite{mueh09b}). Likewise, the A-phase for $\langle100\rangle$ shows two domain populations, thus indicating a flat potential landscape, which may also explain the peculiar regime of the A-phase for $\langle 100\rangle$ (Fig.\,\ref{figure3}\,C). Further, the metastable features of the A-phase are similar to superconducting flux lattices in the presence of disorder. For instance, in the presence of disorder the superconducting flux lattice frequently leads to a ring of intensity in SANS \cite{cubi07}, reminiscent of the A-phase after ZFC. Similarly, for FC the superconducting flux lattices survives as a metastable state in the Meissner phase. This is akin to the field range in which the A-phase during FC is observed down to the lowest temperatures (we attribute the temperature dependence well below $T_c$ to that of the ordered moment). 

An outstanding puzzle is the intensity distribution of the ZFC state, which is a priori not connected with the skyrmion lattice in the A-phase. However, two scenarios have been suggested with a similar context: a metastable body centered multi-$Q$ state \cite{binz06}, or weakly stratified skyrmions lines \cite{roes06}. Thus the ZFC state and its metastable nature raise two exciting issues for future studies. First, whether disorder enhances the stability of skyrmion lattices in an applied field and, second, whether skyrmions may even form spontaneously in zero field.

In conclusion, our small angle neutron scattering study of {\fecosi} identifies a second material after MnSi in which a skyrmion lattice forms. As its most important facet, the skyrmion lattice appears in a strongly doped semiconductor, where disorder leads to hysteretic behavior and the observation of a skyrmion lattice in a wide temperature range.

We wish to thank A. Bauer, B. Binz, F. Birkelbach, S. Dunsiger, M. Garst, S. Legl, M. Kartsovnik, R. Ritz, R. Schikowski, B. Russ, M. Vojta, J. Zweck and W. Zwerger. Financial support through SFB608 of the DFG and NSF grant PHY05-51164 are gratefully acknowledged.


\begin{thebibliography}{32}
\bibitem[{rot()}]{rotate}
\bibinfo{note}{Intensity maxima for $\langle110\rangle$ are the only way to
  to account for the angles of rotation between Fig.\,\ref{figure1}A, B and C.}
\bibitem[{zfc()}]{zfc}
\bibinfo{note}{The orientation dependence is not shown in \cite{ishi95,grig07a,take09}.}
\end{thebibliography}

\end{document}